\documentclass[aoas,nameyear,seceqn,dvips]{arximspdf}
\usepackage{dcolumn}
\usepackage{graphicx}

\doi{10.1214/09-AOAS295}
\volume{4}
\issue{2}
\pubyear{2010}
\firstpage{805}
\lastpage{829}

\makeatletter
\newcolumntype{e}[1]{D{.}{.\ }{#1}}
\makeatother

\begin{document}
\begin{frontmatter}

\title{Changing approaches of prosecutors towards juvenile repeated
sex-offenders:\\ A Bayesian evaluation}
\runtitle{Bayesian analysis of sex-offenders}

\begin{aug}
\author[A]{\fnms{Dipankar} \snm{Bandyopadhyay}\corref{}\thanksref{t1,t2}\ead[label=e1]{bandyopd@musc.edu}},
\author[B]{\fnms{Debajyoti} \snm{Sinha}\ead[label=e2]{sinhad@stat.fsu.edu}},
\author[C]{\fnms{Stuart}~\snm{Lipsitz}\ead[label=e3]{slipsitz@partners.org}}
\and
\author[D]{\fnms{Elizabeth} \snm{Letourneau}\thanksref{t1}\ead[label=e4]{letourej@musc.edu}}
\thankstext{t1}{Supported by CDC Grant R49-000-567.}
\thankstext{t2}{Supported in part by NIH Grants P20 RR017696-06 and
5U10 DA013727-09.}
\runauthor{Bandyopadhyay, Sinha, Lipsitz and Letourneau}
\affiliation{Medical University of South Carolina,
Florida State University,\break
Brigham and Women's Hospital and
Medical University of South Carolina}

\address[A]{D. Bandyopadhyay\\
Division of Biostatistics and Epidemiology\\
Department of Medicine\\
Medical University of South Carolina\\
Charleston, South Carolina 29425\\
USA\\ \printead{e1}}

\address[B]{D. Sinha\\
Department of Statistics\\
Florida State University\\
Tallahassee, Florida 32306\\
USA\\ \printead{e2}}

\address[C]{S. Lipsitz\\
Brigham and Women's Hospital\\
Boston, Massachusetts 02115\\
USA\\ \printead{e3}}

\address[D]{E. Letourneau\\
Family Service Research Center\\
Medical University of South Carolina\\
Charleston, South Carolina 29401\\
USA\\
\printead{e4}}
\end{aug}

\received{\smonth{1} \syear{2009}}
\revised{\smonth{9} \syear{2009}}

%
\begin{abstract}
Existing state-wide data bases on prosecutors' decisions about
juvenile offenders are important, yet often un-explored resources
for understanding changes in patterns of judicial decisions over
time. We investigate the extent and nature of change in judicial
behavior toward juveniles following the enactment of a new set of
mandatory registration policies between 1992 and 1996 via analyzing
the data on prosecutors' decisions of moving forward for youths
repeatedly charged with sexual violence in South Carolina. To
analyze this longitudinal binary data, we use a random effects
logistic regression model via incorporating an unknown change-point
year. For convenient physical interpretation, our models allow the
proportional odds interpretation of effects of the explanatory
variables and the change-point year with and without conditioning on
the youth-specific random effects. As a consequence, the effects of
the unknown change-point year and other factors can be interpreted
as changes in both within youth and population averaged odds of
moving forward. Using a Bayesian paradigm, we consider various prior
opinions about the unknown year of the change in the pattern of
prosecutors' decision. Based on the available data, we make
posteriori conclusions about whether a change-point has occurred
between 1992 and 1996 (inclusive), evaluate the degree of confidence
about the year of change-point, estimate the magnitude of the
effects of the change-point and other factors, and investigate other
provocative questions about patterns of prosecutors' decisions over
time.
\end{abstract}

%
\end{frontmatter}

\section{Introduction}

In the United States, juvenile sex offenders are increasingly being
treated as adult offenders, and are being subjected to similar
punishments and restrictions [Letourneau and Miner (\citeyear
{LetourneauMiner2005})]. In
particular, federal and state sex offender registration and public
notification requirements have been extended now to include
juveniles, with little consideration of differences between adult
and juvenile development or culpability [Garfinkle (\citeyear
{Garfinkle2003})]. Ideally,
these considerations should influence legal responses to juvenile
criminal behavior [Trivits and Reppucci (\citeyear
{TrivitsReppucci2002}); Zimring (\citeyear{Zimring2004})].
A series of landmark registration laws and policies dealing with
juvenile sex offenders was implemented across the US throughout
the mid-to-late 1990s [Chaiken (\citeyear{Chaiken1998})]. Initially,
these important
policies required probationers and paroled sexual offenders to
register personal information with law enforcement (`registration'
laws), but subsequent major amendments increased public access to
registry data (`notification' laws).

These policies, including community notification or public
registration, were enacted with great hope to improve community
safety via either preventing or detecting early recidivism of sexual
offenses (e.g., community members can notify the police about
suspicious behavior by a known sex offender), as well as deterring
sex offenders from committing new sexual offenses (e.g., offenders
may be discouraged from committing new offenses if they believe that
both the police and community members are providing additional
surveillance of their activities) [\mbox{LaFond} (\citeyear{LaFond2005});
Terry and Furlong (\citeyear{TerryFurlong2004})]. The extent of such
intended beneficial effects of these
registration policies has been questioned [\mbox{LaFond} (\citeyear
{LaFond2005})] and
substantive concerns have been raised regarding possible latent
negative consequences of such policies [e.g.,
Edwards and Hensley (\citeyear{EdwardsHensley2001});
Tewksbury (\citeyear{Tewksbury2005});
Zevitz (\citeyear{Zevitz2006})], particularly with respect to
the application of these policies to juvenile offenders
[Trivits and Reppucci (\citeyear{TrivitsReppucci2002});
Zimring (\citeyear{Zimring2004});
Chaffin (\citeyear{Chaffin2008})].

While most studies of latent consequences of such policies have
focused on barriers to the successful reintegration of offenders
into society [Levenson and Cotter (\citeyear{LevensonCotter2005})], it
also has been theorized
by some that the perceived severity of some registration policies
might have the unintended effect of reducing the likelihood of
formal prosecution [\mbox{LaFond} (\citeyear{LaFond2005})]. One recent survey
of family and
juvenile court judges reported that the majority of judges believed
`registration' could be harmful to juvenile offenders
[Bumby, Talbot and Carter (\citeyear{BumbyTalbotCarter2008})].
Arguably, such a perception of these policies may affect the
decisions of prosecutors and judges. We would like to evaluate the
strength of the available data evidence to support the hypothesis of
judicial decision makers (particularly prosecutors) becoming less
likely to prosecute a juvenile charged with sexual offense during
the period of enactment of these policies. To address this, we
analyze the data on prosecutors' decision making pattern toward
juvenile sexual offense charges during 1988--2005, a period of
time encompassing the enactment of sex offender registry and other
related laws/amendments in South Carolina (SC). To further
understand the extent of the change in decision patterns of the
prosecutors, we would also like to estimate the magnitude of this
change (if it exists) in terms of change in odds of prosecution
after the change-point time and determine other factors affecting
the prosecutors' decisions during this period.

\subsection{Prosecution of juvenile sex offenders}

We now explain the reasons for focusing on responses from
prosecutors for understanding the changes in the judicial decision
makers' actions toward juvenile offenders. After a certain charge
has been brought against a youth by law enforcement, the prosecutor
is the first judicial decision maker encountered by the youth. Other
judicial decision makers such as judges play their respective role
only after a prosecutor's decision to move forward on the case has
been made. South Carolina has a very well maintained juvenile
justice database which serves as an invaluable resource for
examining the change in patterns of prosecutor's decision-making. We
reiterate that we decided to use the prosecutors' decisions to \textit
{move forward} on initial felony sexual offense cases as the response
variable relevant to prosecutors decisions.

Like many other states, SC's registration and notification policies
enacted in 1995 and 1999, respectively (SC Code of Laws $\S\S$
23-3-400 et seq.) exceed, in every respect, the original federal
registration and notification requirements established in the 1990s
(e.g., Federal 42 USCA $\S$ 14071, 14072; Pub. L. No. 104-145,
110 Stat. 1345) and continue to exceed even the expanded
requirements more recently established by the Adam Walsh Child
Safety and Protection Act of 2006. To give some examples of the
deviations from federal guidelines, SC's policies (a)~require
lifetime registration with no exceptions; (b) have no lower age
limit for the registration of sexual offenders (e.g., children as
young as 10 years of age have been required to register) and make
few distinctions between juvenile and adult offenders; (c) make no
distinctions between low and high risk offenders; and (d)~include
noncontact sexual offenses such as voyeurism and indecent exposure.
Due to the breadth and great severity of SC's policies and their
application to teens and pre-teens, it is anticipated that SC's
policies have unintended effects of reducing the probability of
prosecutor's decision of moving forward with the initial sexual
charge.

\subsection{Focus on repeat sexual offense charges}

We have chosen the binary response of prosecutor decision to
prosecute (called `moving forward') as the primary response variable
related to judicial decision. We investigate the nature and
magnitude of changes in probability of moving forward with juvenile
sexual offense by studying the available data about an interesting
group of 358 male youths charged with sexual offenses in SC at least
twice. The effects of public registration policies on repeat
offenders have not yet been examined (for both adult or juvenile
offenders) but they are relevant for several reasons. First,
registration and notification policies were developed in response to
public outrage over especially heinous and well-publicized sexual
offenses by known sex offenders [\mbox{LaFond} (\citeyear{LaFond2005})].
These policies were
intended to target only the most severe offenders who were unlikely
to be juveniles because sexual recidivism is rare for youth
offenders [Fortune and Lambie (\citeyear{FortuneLambie2006})].
However, a repeat offense
charge might serve as a useful indicator of offender severity.
Second, as has been demonstrated in previous research [Howell
(\citeyear{Howell2003})],
number of prior offenses (any offense, not just sexual) may
influence the likelihood of cases moving forward. There was some
evidence to suggest that prosecutors may distinguish between
first-time and repeat offenders. Whether this finding applies to
repeat juvenile sex offenders is unknown.
Third, if it is found that
the risk of registration negatively influences the likelihood of
prosecution even for repeat offenders [as it apparently does for
first-time offenders; see Letourneau et al. (\citeyear
{LetourneauEtAl2009})], the public
policy implications of such a finding would be profound. Juvenile
justice encounters for sex crimes contribute little to the
prediction of adult sex crimes [Caldwell (\citeyear{Caldwell2002});
Zimring, Piquero and Jennings (\citeyear
{ZimringPiqueroJennings2007})]. Nevertheless, repeat juvenile sex
offenders represent a small
but important subgroup that might benefit from additional
surveillance and treatment that accompanies adjudications. Most
sexual offenders in the US receive treatment only after legal
prosecution and undetected sex offenders have little motivation to
seek out professional help. Thus, youths who are not prosecuted for
serious sexual offenses are unlikely to receive proper treatment or
supervision. Recent evidence of sex offender treatment effectiveness
[Letourneau et al. (2009)] makes evidence-based treatment of these
youth all the more compelling.

There can be substantial amount of heterogeneity among youth
offenders due to how differently they interact with judiciary. This
unobservable interaction for each youth will be modeled by a
youth-specific random effect. Analyzing data with repeated ($\geq$2
per youth) binary responses of prosecutors' decisions of moving
forward at different time points will allow us to assess the effects
of the unobservable prosecutor-youth interaction on prosecutor's
decision. Available data will facilitate the assessment of whether
prosecutors treat youths differently during a first offense charge
compared to a subsequent charge. Thus, following the change-point
year, prosecutors might become more lenient on first time offenders,
while at the same time they might adopt a more hardened approach to
recidivists.

%
%
\begin{figure}

\includegraphics{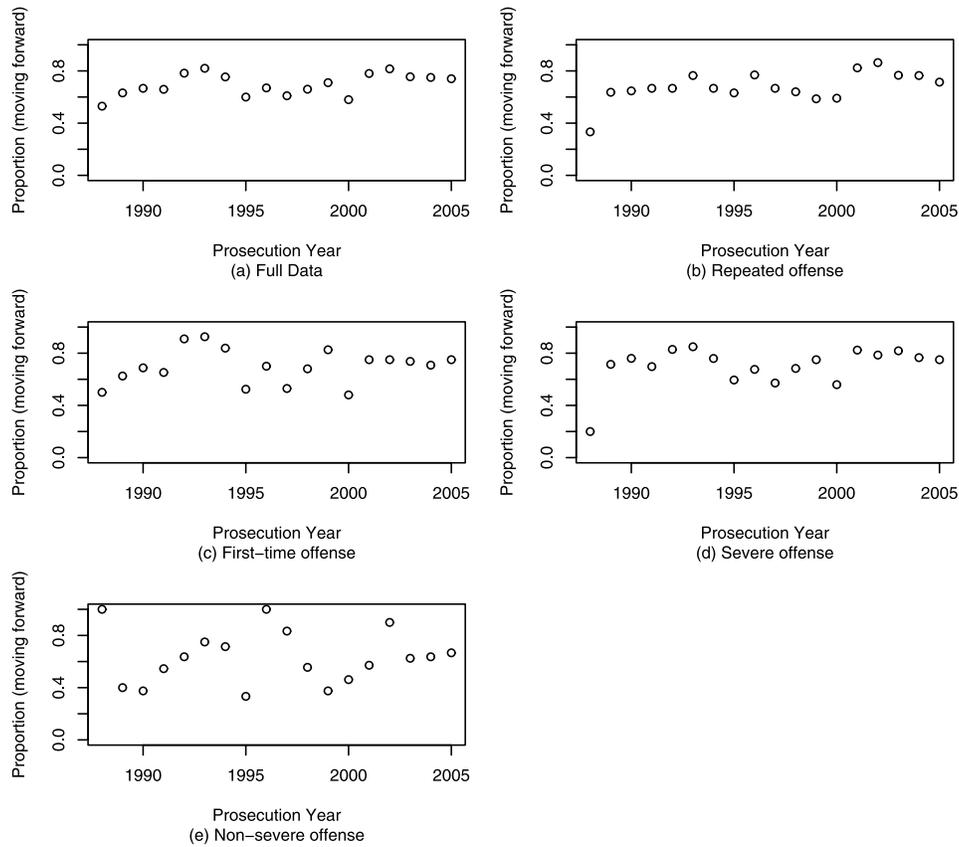}

\caption{Panel plots of observed proportion $p$ of prosecutor's
moving forward versus prosecution years.
Panel \textup{(a)}: full data,
panel \textup{(b)}: repeated offense,
panel \textup{(c)}: first-time offense,
panel \textup{(d)}: severe offense and
panel \textup{(e)}: nonsevere offense.}\label{FIG1}
\end{figure}

\subsection{Unknown change-point}

As noted before, our available data spans from 1988 through 2005. It
is conceivable that changes in patterns of prosecution could occur
in any one (or more) of these years. Given implementation of South
Carolina's registration policy in 1995, we are most interested in
determining whether there is substantial data evidence for 1995 as a
change-point year. As seen in the plot of the raw data [Figure~\ref{FIG1}(a)],
there was a high magnitude of decrease in the observed
proportion of prosecutors moving forward from 1994 to 1995 and this
decrease appears substantial relative to other fluctuations
occurring within the full time interval. However, it is difficult
and naive to make any conclusion about possible change-point year
and magnitude of change in probability of moving forward from these
raw proportions (computed via ignoring effect of any covariate).
Furthermore, state data on juvenile arrest rates (for rape) also
confirm a substantial drop around 1994--1995 [McManus (\citeyear
{McManus2005}), page~161],
suggesting the influence of the 1995 legislation on other judicial
actors. Last, our own previous research supports a change in
prosecution patterns in 1995 for a data set with predominantly single
offense charges [Letourneau et al. (\citeyear{LetourneauEtAl2009})].

There has been almost no research on repeat offenses by juvenile sex
offenders and, thus, it is an open argument as to whether the possible
change-point year for this smaller subset of offenders should be
restricted to a single year, 1995 in particular. The mandatory
registration policy enacted in South Carolina in 1995 might not be
the main landmark policy for this group. For example, 1996 was the
year when several sex crimes were classified as `no parole' crimes
as part of South Carolina's `truth in sentencing' policies.
Likewise, changes in juvenile transfer policies that made it easier
to transfer younger defendants (14--15 years old) to adult court for
certain offenses also occurred in 1996. Thus, these or other policy
or policy changes (e.g., lengthened sentences) could exert influence
on the pattern of juvenile judicial decision making. In spite of
using our apriori belief that the most likely year for
change is 1995, our discrete prior distribution of the unknown
change-point year in between 1992 and 1996 will reflect a skeptical view
that laws enacted in years other than 1995 might also have caused
the change in pattern of prosecutors' decisions. It could also be
argued that legal policies take time to reach their full effect on
decision makers (e.g., reflecting a learning curve among judicial
decision makers regarding the severity of a policy) and that
focusing on a single year of sudden change-point is unnecessarily
narrow. To address this concern, we incorporate a linear and a
quadratic coefficient of change-point effect to provide the
flexibility of the nature of effects generated by the change-point
(if it exits).

If we find credible data evidence for any change in prosecutor's
pattern of moving forward during 1992--1996, we would further like to
find out the most likely year of change-point as well as the
magnitude of the change in terms of odds ratio before and after the
change-point year. The actual magnitude of the change-point effect
is important to understand the practical consequences to society
for such a change. We will handle these complex goals using a model
for longitudinal binary data where the change in the odds of going
forward after the change-point year can be estimated. Our model is
very different from the unknown change-point model used for single
time-series of quantitative responses discussed in Carlin and Louis
(\citeyear{CarlinLouis2000}) and other related works.

\subsection{Youth heterogeneity and conditional/marginal odds ratio}

Large sample based analysis using generalized estimating equations
(GEE) for multivariate clustered binary outcomes
[Zeger and Liang (\citeyear{ZegerLiang1986})] is very common in
biomedical and behavioral studies. However,
the unobservable prosecutors' reaction to different individual
youths (youth specific youth-prosecutors interaction effect) can
substantially attenuate the actual subject-specific effects of
covariates and time/year and change-point on the pattern of judicial
decision making. The marginal GEE approach is unable to estimate
this degree of attenuation. This attenuation is a measure of
variability among youths regarding some youths being more
predisposed (compared to others) to receive a moving forward decision.

In our analysis, we express this latent/unobservable youth-specific
interaction as a random youth effect. We are interested in
determining the amount of attenuation in the effects of the
change-point year and would like to make probability statements
about the actual change in probability of moving forward in a
particular year (say, 1995) based on observed data. Consequently, we
want to maintain key advantages of a marginal model based GEE
approach such as simple physical interpretation of covariates and
change-point year effects in terms of the marginal odds ratio. To
achieve this goal, we present an extension of the random intercept
model proposed by Wang and Louis (\citeyear{WangLouis2003}) in which the
subject-specific model (conditional on the youth-specific random
intercept) as well as the marginal model (integrated over the
distribution of the unobservable random intercept) have the same
link functions. The regression parameters in our conditional and
marginal models are not identical, but are proportional to each
other and the proportionality parameter will represent the
attenuation of all the covariate and change-point effects on
marginal response due to heterogeneity of youths. For example,
consider a youth with two offenses committed before and after the
change-point year. Our model will be able to assess the
individualized odds of moving forward for the offense before
change-point year and the odds of moving forward for the same
youth's offense committed after the change-point year. The model
will also be able to assess the marginal odds ratio of moving
forward for two offenses committed by two different youths. This
marginal odds ratio is smaller than the corresponding
conditional/individualized odds ratio.

\section{Brief overview of the data}\label{sec2}

Every South Carolina youth (male) charged by the Department of
Juvenile Justice (DJJ) with repeated felony-level sexual offenses
between January 1, 1988 and December 31, 2005 is included for the
present analysis. Female offenders were excluded due to the small
number of females in the database. Charges filed against minors in
general sessions (adult court) were not included in the study. The
analysis focused on felony sexual offenses because South Carolina's
registration and notification policies primarily apply to sexual
offenses. All data were drawn from the South Carolina DJJ Management
Information System in collaboration with the South Carolina Budget
and Control Board Office of Research and Statistics. The data
included a subset of variables regularly captured by DJJ personnel
during processing of each charge and subsequently during progression
of the case, viz. (a) demographic information, (b) youth's criminal
history, and (c) information regarding specific sexual felony charge
(viz. type of charge, degree of severity, charge date, prosecutor's
decision, etc.).

The database included records for 358 male offenders with a total of
753 offenses. For each charge, the prosecutor decision could
indicate moving forward (i.e., decision to formally adjudicate the
youth), diversion (i.e., decision to refer youth to a nonjudicial
intervention that, if successfully completed, would clear the initial
charge), or dismissal (i.e., decision to not process or otherwise
dismiss~the~case). The longitudinal response variable $Y_{it}$ for
youth $i$ at calendar time~$t$ is the binary indicator with
$Y_{it}=1$ for moving forward (70.7\% of cases) and $Y_{it}=0$ for
diversion or dismissal of the charge by the prosecutors (29.3\% of
cases).

For each youth, a set of known/observable explanatory variables
(fixed or time-dependent) were recorded that may influence the
probability of prosecutor's going forward. The set of fixed
explanatory variables include age at the time $t$ of the offense, an
indicator of repeated offense, severity index of the offense,
prosecution year, etc. The median age for this group of male youths
was 14.6 years with a range of 9--19 years. The charge severity
rating for each felony sexual charge determined by DJJ is based on
the number of years an adult would be incarcerated for a similar
crime. The charge severity ratings range from 1 (lowest level
misdemeanor) to 25 (highest level felony and typically reserved for
1st degree murder charges), with felony offenses operationally
defined as charges with severity ratings of 5 or higher
[Barrett, Katsiyanis and Zhang (\citeyear
{BarrettKatsiyanisZhang2006})]. In practice, felony level sex crimes
had severity
ratings of 5, 8, 15 and 21. For simplicity, we use a binary
indicator to classify the severity rating as severe (80.5\% cases
with severity rating $>$ 8) and nonsevere (19.5\% cases with
severity rating $\leq$ 8). Similarly,
we use a binary indicator to
record whether a particular offense is a repeated offense. Out of
the total 753 felony level sex offenses, 395 (52.4\%) were repeated
offenses and the rest 358 (47.6\%) were the first-time offenses.
Table~\ref{TABLE1} shows the $2\times2$ table for the actual counts
and proportions
for charges with two severity levels versus first-time/repeat
charges.
%
\begin{table}
\caption{$2\times2$ table showing frequencies of repeat/first-time
offenses vs. severe/nonsevere offenses} \label{TABLE1}
\begin{tabular*}{\textwidth}{@{\extracolsep{\fill}}lccc@{}}
\hline
& \textbf{Severe offense} & \textbf{Nonsevere offense} & \textbf{Total} \\
\hline
Repeated offense & 284 (38\%) & 111 (15\%) & 395 \\
First-time offense & 322 (43\%) & \phantom{0}36 (4\%)\phantom{0} & 358 \\[6pt]
Total & 606\phantom{ (43\%)} & 147\phantom{ (43\%)} & 753 \\
\hline
\end{tabular*}
\end{table}
There were 481 cases (63.8\%) of prosecution conducted after the
registration policy was implemented in January 1995. Of the 358 male
offenders, 326 (91.1\% cases) had 2 offenses. Among the remaining 32
youth offenders, 28 (7.8\% cases) had 3 offenses, 3 (0.83\% cases)
had 4 offenses and 1 (0.27\% cases) had 5 offenses. Thus, each
offender represents a cluster with the maximum cluster size being 5.
Figure~\ref{FIG1} shows panel plots of the raw proportion of prosecutors
moving forward vs. year of prosecution for all charges (panel~a),
charges for repeated offenses (panel~b), charges for first-time
offenses (panel~c), charges for severe offenses (panel~d) as well as
charges for nonsevere offenses (panel~e).
The figure suggests some possible effect around 1995; however, the
actual year of change-point and its association with the type of
charge (severe/first-time/repeat) are not clear from the plots. The
evidence of a latent change-point between 1992 and 1996 and its
magnitude of influence on prosecutors' decision and other
conjectures about the pattern of decisions over time can only be
evaluated with a semi-continuous (change-point) model along with
linear and quadratic effects of change-point (that determines
whether this change effect was gradual over years). Any interaction
of each known explanatory covariate with the change-point indicator
(viz. repeat offense indicator with change-point indicator) will be
considered as a time-dependent covariate. In the year 1999, the
sexual offender registry became available online and about half of
registered juveniles were included. In addition to the unknown
change-point effect, we will also attempt to verify whether the
notification law of 1999 influenced the prosecutor's decision since
the year 2000. There are 252 cases (33.5\%) of prosecution after the
implementation of the online registry in 1999. Out of our concern
for statistical association between different pairs of explanatory
variables (after properly accounting for clustering), we used the
non-Bayesian Rao--Scott chi-square test (available in SAS Procedure
SURVEYFREQ) to find strong evidence of association between the
repeat offense indicator and the severity indicator ($p$-value $<$
0.001). We did not expect any association among all other variables
and similar frequentist tests for evaluating association between the
remaining explanatory variables are not statistically significant.
In our formal Bayesian analysis, we will evaluate whether both
severity indicator and repeat offense indicator should be included
simultaneously as predictors.

The time-interval of the study (1988--2005) permitted examination of
cases processed during years prior to implementation of the South
Carolina sex offender registration (i.e., January 1, 1988--December 31,
1994), as well as cases processed during the four years
following registration but prior to implementation of the online
registry website (i.e., January 1, 1995--December 31, 1998)
and cases processed during the years subsequent to implementation of
the Internet-based public registration (i.e., January 1, 1999--December
31, 2005). One of our aims in this article is to draw
statisticians' attention to this important class of databases. These
data can be extensively and critically modeled and analyzed to
investigate whether our judicial decision making process is changing
with time, with ever-shifting societal perception and evolving
legislative activism. Assessing the existence and effect of the
change-point time induced by mandatory registration laws on the
prosecutor's decision of moving forward with a juvenile sex offense
case has tremendous societal implications in its own right and hence
demands serious exploration of available evidence via these
databases.

\section{Bridge random effects model}

Our modeling goal is to interpret the effects of change-point and
other factors on the changes in `odds of moving forward' for a
particular charge. The following model using a particular random
effects density (called the `Bridge' density) preserves the odds
ratio interpretation of the change-point and other factors.

For $i=1,\ldots,n$ exchangeable youths/subjects, sexual charges were
brought against youth $i$ at years $t=t_{i1},\ldots,t_{im_i}$ for
$m_i\ge2$. The binary response of interest is $Y_{it}=1$ if the
prosecutor decides to move forward (proceed with the prosecution)
for the sexual offense charge against the youth $i$ at time $t$, and
$Y_{it} = 0$ otherwise. For the random effects logistic regression
model, the conditional probability $p_{it}(B_i)=\operatorname
{pr}[Y_{it}=1|B_i,\underline{x_{it}}]$, given the subject-specific
unobservable $B_i$ and the known $p \times1$ covariate vector
$\underline{x_{it}}$ measured at time $t$, is
%
\begin{equation}
p_{it} = \frac{\exp( B_i +
\bolds{\beta}\underline{x_{it}'})}{1 + \exp( B_i +
\bolds{\beta}\underline{x_{it}'})}, \label{cond}
\end{equation}
\noindent where $\bolds{\beta}=(\beta_1,\ldots,\beta_p)$
denotes the vector of regression parameters. The offender-specific
unobservable random effects $B_i$ has density $f_B(b|\phi)$ with the
variability parameter $\phi$. The binary model with logit link in
({\ref{cond}}) gives an easily understandable proportional odds
interpretation of the covariate effects when we know the
unobservable youth-specific $B_i$. However, the change in odds ratio
interpretation of the regression effects is not preserved in general
after integrating out $B_i$. When $B_i$ follows a bridge
distribution of Wang and Louis (\citeyear{WangLouis2003}) with density,
%
\begin{equation}\label{bridge}
f_B(b|\phi) = \frac{1}{2\pi} \frac{\sin(\phi\pi)}{\cosh(\phi b) + \cos(\phi\pi)},\qquad
-\infty< b < \infty,
\end{equation}
indexed by unknown parameter $0 < \phi< 1$, the marginal
probability of moving forward (after integrating unobservable
youth-specific $B_i$), still preserves the logistic form unlike
normal and $t$-density random effects and is given as
%
\begin{equation}\label{marg}
\operatorname{Pr}( Y_{it}=1|\underline{x_{it}}) = E_B [ p_{it}(B_i) ]
=\frac{\exp[ (\phi\bolds{\beta})\underline{x_{it}'}]}{1 +
\exp[(\phi\bolds{\beta})\underline{x_{it}'}]},
\end{equation}
where $E_B$ denotes the expectation with respect to the density of
$B_i$.
%
\begin{figure}[b]

\includegraphics{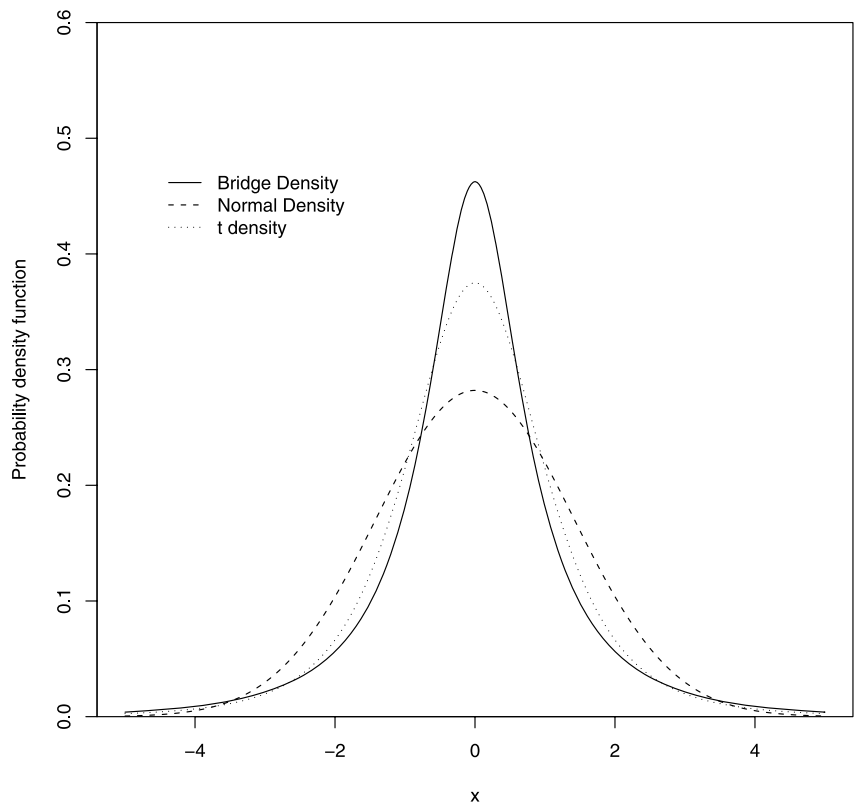}

\caption{Probability density functions of Bridge, Normal and
$t$-densities with zero mean and variance $=$2. The $x$-axis denotes the
range of $x$-values from $-$5 to 5.} \label{FIG2}
\end{figure}
The bridge density is symmetric around mean zero [Wang and Louis (\citeyear{WangLouis2003})] with the variance given by $\sigma_B^{2}=
\pi^2(\phi^{-2}-1)/3$. Figure~\ref{FIG2} displays the comparison of the
bridge density (with variance $=$ 2) with a $\operatorname{Normal}(0,\sigma^2=2)$ and
a $t$-density having degrees of freedom $\nu=4$ (corresponding to a
variance of 2). The bridge density has a slightly heavier tail and
is more peaked than the normal and the $t$-densities. We again
emphasize that, unlike the normal and the $t$-density, both the
conditional probability in ({\ref{cond}}) and the marginal probability
in ({\ref{marg}}) of moving forward under the bridge density
random-effects have logistic links, with proportional odds
interpretations of the regression effects.
%
\begin{figure}[b]
\vspace*{3pt}

\includegraphics{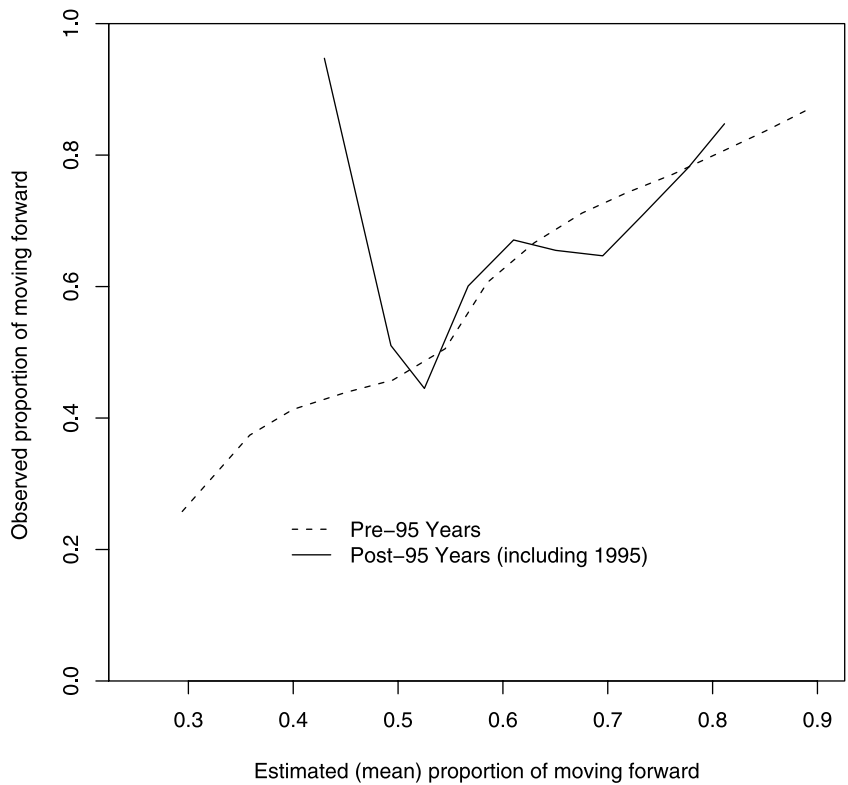}

\caption{LOWESS smoothed plots of observed proportion of moving
forward vs. estimated (mean) proportion of moving forward for pre-95
and post-95 years.} \label{FIG3}
\end{figure}
To assess the need of incorporating a change-point structure into
our logistic regression framework, we divided the whole data into
two sections, (i) prosecution year $<$ 1995 and
(ii)~prosecution year $\ge$ 1995, and fit separate logistic regression models (using
PROC LOGISTIC) in SAS, considering all data points to be independent
in each of the models. We used prosecution age, severity indicator,
repeat offense indicator and prosecution year as covariates without
any change-point term. Then, the observed proportion of prosecutors
`moving forward' versus the estimated (mean) proportion of
prosecutors `moving forward' (after LOWESS smoothing) were plotted
in Figure~\ref{FIG3}, overlayed on each other. The simple logistic model for
the `post-95' data is clearly inadequate, as expected, due to the
effect of the change-point somewhere around 1995. Clearly, there is
a need to account for the (possibly unknown) change-point structure
in our model. The Bayesian paradigm allows us to use effectively our
apriori belief about the occurrence of the change-point.

In this article we will use a novel extension of the longitudinal
binary model of ({\ref{cond}}) by allowing time-dependent covariates
$\underline{x_{it}}(T)$ which are functions of unknown year of
change-point $T$ as well as calendar time $t$. To incorporate the
effects of an unknown change-point year $(1992\le T\le1996)$, the
covariate vector has two components, viz. (a) a vector
$\underline{x_{1it}}$ of either fixed or time-dependent
known/recorded covariates such as age at charge, binary indicator of
repeat offense for the charge, a dichotomized indicator for severity
of offense, an indicator for the year of notification `2000,' the
prosecution year $t$, and (b) a vector $\underline{x_{2i}}(t,T)$ of
known functions of prosecution year $t$ and unknown change-point
year $T$ common to all subjects. The functions of unknown $T$
include the change-point indicator $1_{[t\ge T]}$, a linear
coefficient of the change-point indicator $(t-T)1_{[t\ge T]}$ and
also a quadratic coefficient $(t-T)^21_{[t\ge T]}$. The
change-point term $1_{[t\ge T]}$ accommodates a sudden change in
the pattern at time $T$. The linear and quadratic terms allow the
change in pattern to be continuous after the unknown change-point
year $T$. We will later discuss how to use discrete prior
distribution of $T$ to reflect our prior opinion about what
value/year $T$ can take in the interval 1992--1996. We also include
the interaction of `repeat offense' and `change-point indicator'
$1_{[t\ge T;\mathit{Repoff}=1]}$ to assess whether the magnitude of the
change in prosecutors' decision pattern after $T$ depends on whether
the charge is a first time offense versus a repeat offense.

Our extension of the model in ({\ref{cond}}) is given as
%
\begin{equation}\label{regr}
\operatorname{logit}\{ p_{it}(B_i,T)\} = B_i +
\bolds{\beta_1}\underline{x'_{1it}}+
\bolds{\beta_2}\underline{x'_{2i}}(t,T),
\end{equation}
where the vectors $\bolds{\beta_1}$ and $\bolds{\beta_2}$
represent respectively the regression parameters associated with the
observable fixed/time-dependent covariates and the effect of the
unknown change-point $T$ at year $t$.
Using our model, we can interpret the effect of each covariate,
because we can obtain the marginal as well as the conditional odds
ratio. For example, if we define the parameter for the indicator of
repeat offense to be $\beta_{\mathit{Repoff}}$, then the conditional odds
ratio $e^{(\beta_{\mathit{Repoff}})}$ represents the ratio of the odd of
moving forward when an individual is charged with a repeat offense
and the odd of moving forward when the same charge is for a first
offense. The attenuated marginal odds ratio $e^{(\phi
\beta_{\mathit{Repoff}})}$ measures the ratio of the odd of moving forward
for a youth charged with repeat offense and the odd for another
similar youth charged with a first offense. The variability
parameter $0<\phi<1$ of the bridge density measures both the degree
of attenuation of the marginal/population effect versus
conditional/individualized effect and the heterogeneity of the
decision making process. The extreme case of $\phi=1$ represents the
situation when there is no effect of the unobservable
youth-prosecutor interaction and the responses corresponding to all
the charges from all the youths can be considered as exchangeable.
The extreme case of $\phi=0$ represents the situation when the
unobservable youth-prosecutor effect is so high that it alone
determines the decision of moving forward. Unless we get strong
data evidence against the bridge density for $f_B(\cdot|\phi)$, we will
prefer using the model defined by ({\ref{cond}}) and ({\ref{bridge}})
because this model ensures the convenient proportional odds
interpretation of the effects of covariates and change-points both
conditionally and marginally, and offers a simple role of the
heterogeneity parameter $\phi$ as the attenuation factor of the
marginal odds ratio of two subjects in the presence of heterogeneity.

Using the regression model of ({\ref{regr}}), we can write the
likelihood of $\bolds{\beta}$, the unknown $T$ and the random
effects $\mathbf{B}=(B_1,\ldots,B_n)$ based on the observed data $\mathbf{y}$ as
%
\begin{equation}
L(\bolds{\beta},\mathbf{B},T|\mathbf{y})\propto
\prod_{i=1}^n\prod_{j=1}^{m_i}
\{p_{it_{ij}}(B_i,T)\}^{y_{it_{ij}}}\{1-p_{it_{ij}}(B_i,T)\}
^{y_{it_{ij}}}.
\end{equation}
To draw conclusions about the effects of covariates and the unknown
change-point year on the odds of moving forward, we used a Bayesian
analysis of the random effects logistic regression model
accommodating unknown change-point time $T$. The posterior
conclusion for a Bayesian analysis is based on the joint posterior
of all the parameters given by
%
\begin{eqnarray}\label{post}
p(\bolds{\beta},\mathbf{B},T,\phi|\mathbf{y})&\propto& L(\bolds{\beta},\mathbf{B},T|\mathbf{y})\nonumber
\\[-8pt]\\[-8pt]
&&{}\times\Biggl[\prod_{i=1}^nf_B(B_i|\phi)\Biggr]\times
\pi_1(T)\times\pi_2(\bolds{\beta}) \times\pi_3(\phi),\nonumber
\end{eqnarray}
where $\pi_1(T)$, $\pi_2(\bolds{\beta})$ and $\pi_3(\phi)$ are
independent priors of $T$, $\bolds{\beta}$ and $\phi$ of the
bridge density $f_B(B_i|\phi)$ of ({\ref{bridge}}). The key advantage
of relying on Bayesian inference to address the pertinent question
of the influence of the sex-offender registration laws on
prosecutor's decision making is the ability to incorporate
background (prior) information about the unknown parameters
including the unknown change-point year $T$. Thus, a proper
selection of prior information is an important step toward making
an informed conclusion about the data evidence from the study.
Unlike the frequentist inference depending on large-sample
inference, the Bayesian method relies heavily on the simulations
from the posterior of ({\ref{post}}) via the Gibbs sampler
[Gelfand and Smith (\citeyear{GelfandSmith1990})] and associated Markov chain Monte Carlo (MCMC)
tools. The MCMC method provides the entire posterior distribution of
any arbitrary functional of the parameters. In the following
section we discuss the determination of practical prior
distributions for $\phi$, discrete time $T$ and regression parameter
$\bolds{\beta}$.

\section{Choice of priors}\label{sec4}

In this section we develop practical informative prior
distributions for the parameters including the attenuation parameter
$\phi$ and the unknown change-point year $T$. While selecting a
prior density class/model, we prefer a class of densities with a
small set of informative features to be selected by
investigators/statisticians. For each component $\beta_k$ for
$k=1,\ldots,p$ of the regression parameter $\bolds{\beta}$, we
use the zero centered double exponential (DE) prior (often called
Laplace prior) $\beta\sim \operatorname{DE}(0,\tau)$, with p.d.f.
$\frac{\tau}{2}\exp(-\tau|\beta|)$ for $ -\infty< \beta< \infty$,
and variance $2/\tau^2$. The DE prior is a widely used sparsity
inducing prior with a heavy tail and peakedness at 0, and hence
expressing the prior belief that the distribution of $\beta_k,
k=1,\ldots,p$ is strongly peaked around 0 [Kaban (\citeyear{Kaban2007})]. This prior
reflects our `skeptical' views about the covariate and change-point
effects. Our strategy is to hold such a skeptical prior view unless
there is a strong data-evidence suggesting otherwise. We choose the
precision parameter $\tau=\sqrt2$ for each $\beta_k$. This choice
of~$\tau$ makes the prior variance of $\beta_k$ to be one. Assuming
$\beta_{k}=1$ and a logistic model with a single covariate, this
corresponds to a prior belief that a change in 1 unit of a covariate
can change the value of probability $p_{it}$ of going forward from
0.5 to 0.73. For this data example, we believe our prior belief
allows for a large enough effect of each covariate (keeping other
covariates fixed) because 23\% change in $p_{it}$ for a unit change
of a covariate is a very large change to expect in practice.

Our first prior for $\phi$ also represents a `skeptical' (however,
informative) opinion that heterogeneity among young offenders does
not cause large attenuation of regression effect in marginal
probability of moving forward. To assure this, a $\operatorname{Beta}(2,1)$ prior
density (with mean 0.67) is chosen for $\phi$ to make the prior
skewed toward 1. As an alternative, a $\operatorname{Beta}(1,1)$ [the $\operatorname{Uniform}(0,1)$
prior] is considered as a competing `noninformative' benchmark prior
for $\phi$.

We believe that the change-point year is within the interval
1992--1996 (if the change-point had taken place at all), with the
most plausible year being 1995.\vspace*{1pt} We use a Dirichlet prior on the
years 1992--1996, that is, $ (\gamma_{92},\ldots,\gamma_{96}) \sim
\operatorname{Dir}(\alpha_{92}, \ldots, \alpha_{96}), $ with joint density
$\pi(\gamma_{92},\ldots,\gamma_{96}) \propto
\prod_{j=92}^{96}\gamma_j^{\alpha_j}$, where $\gamma_j =
\mathrm{P}[T=j]$ and prior expectation of $\gamma_j$ equals
$\alpha_j/\alpha_+$ with $\alpha_+=\sum_{j=92}^{96}\alpha_j$. We
provide more prior weight to 1995 (landmark registration year) being
the unknown $T$ compared to the other years. A prior belief very
enthusiastically favoring 1995 as~$T$ is given by
(a)~$(\alpha_{92},\ldots,\alpha_{96})=(1,1,1,6,1)$. This prior implies
the expected prior probability that the change-point happened in
1995 is 0.6, and the rest of the years in 1992--1996 has equal prior
probability. To represent a more skeptical belief, we consider two
more choices of the Dirichlet parameters, viz.
(b)~(1.6, 1.6, 1.6, 5, 1.6) representing an expected prior probability of 0.45 for
$T=1995$ and 0.14 for each of the other years and (c) (1.5, 1.5, 1.5, 3, 1.5) representing an expected prior probability of 0.33 for
$T=1995$ and 0.167 for the other years. It is important to note that
the prior in~(b) has higher $\alpha_{+}$ than~(a) and~(c). This
implies that the certainty about the value of each $\gamma_j$ is
higher in~(b) compared to the other priors.

For a better description of the practical implication of our prior
densities, we also determine the extent of the effect of each
component of $x_{it}$ on the `change in odds of moving forward' as
expected by our prior choice. We recall that the marginal prior
density of each component $\beta_k$ is assumed to be identical here.
Based on two prior choices of $\phi$, that is, $\phi\sim$
$\operatorname{Uniform}(0,1)$ and $\phi\sim \operatorname{Beta}(2,1)$, the change in odds of
moving forward can be anywhere between 0.05 and 14.4 with 95\%
probability. This shows that our particular joint priors allow for
the possibility of a very large change in odds of moving forward in
either direction. To address a practical concern about whether each
of our prior models allow wide flexibility about the probability of
moving forward at any particular year, we verified whether the range
of possible values of the probability
%
\begin{equation}
p^*_{T,t}=\frac{\exp(\phi\bolds{\beta}\underline{x^*}'(t,T))}{1
+ \exp(\phi\bolds{\beta}\underline{x^*}'(t,T))}
\end{equation}
are wide enough for all possible values of $t$ and $T$. We consider
$p^*_{T,t}$ as the probability of moving forward for a randomly
selected youth (with median age) charged with a severe repeat sexual
offense in year $t$ when the change-point year $T$ is known. For
$t=1996$ and $T=1995$, the range between 0.004 and 0.997 indicates
that our prior belief allows a wide range of possible values of the
probability of moving forward at any year.

\section{Bayesian computation and model selection}

In this section we discuss the MCMC computation and model selection
and assessment procedures for eight competing models, determined by
the sets of prior assumptions for change-point $T$ and the
attenuation parameter $\phi$. The reader less interested about
technical details about model selection and MCMC computation can
skip this section and proceed to the next section for detailed
analysis and findings from the study. The competing models under
consideration are as follows:
\begin{longlist}
\item[\textit{Model}-1:] $(\gamma_{92},\ldots,\gamma_{96}) \sim$
Dir$(1,1,1,6,1)$ and $\phi\sim \operatorname{Uniform}(0,1)$;

\item[\textit{Model}-2:] $(\gamma_{92},\ldots,\gamma_{96}) \sim$
Dir$(1.6,1.6,1.6,5,1.6)$ and $\phi\sim \operatorname{Uniform}(0,1)$;

\item[\textit{Model}-3:] $(\gamma_{92},\ldots,\gamma_{96}) \sim$
Dir$(1.5,1.5,1.5,3,1.5)$ and $\phi\sim \operatorname{Uniform}(0,1)$;

\item[\textit{Model}-4:] $(\gamma_{92},\ldots,\gamma_{96}) \sim$
Dir$(1,1,1,6,1)$ and $\phi\sim \operatorname{Beta}(2,1)$;

\item[\textit{Model}-5:] $(\gamma_{92},\ldots,\gamma_{96}) \sim$
Dir$(1.6,1.6,1.6,5,1.6)$ and $\phi\sim \operatorname{Beta}(2,1)$;

\item[\textit{Model}-6:] $(\gamma_{92},\ldots,\gamma_{96}) \sim$
Dir$(1.5,1.5,1.5,3,1.5)$ and $\phi\sim \operatorname{Beta}(2,1)$;

\item[\textit{Model}-7:] model with no change-point;

\item[\textit{Model}-8:] model with the change-point year fixed at 1995.
\end{longlist}

The prior for the regression parameters is the same for each of the
above prior models. For each prior model representing different
prior opinion, the computation of the posterior distribution was
performed via iterative MCMC [Gelfand and Smith (\citeyear{GelfandSmith1990})] techniques.
To sample from the joint posterior under each prior model, we need
to sample from the conditional distributions of different model
parameters. Each conditional posterior is proportional to the joint
posterior density (6) as a function of the corresponding parameter.
The conditional posteriors of $B_i, \bolds{\beta}$ and $\phi$
are log-concave (proofs omitted), therefore, standard adaptive
rejection algorithms of Gilks and Wild (\citeyear{GilksWild1992}) are applicable to
sample from these conditional posteriors. The relevant Gibbs sampling
steps were implemented readily using freeware package $\tt{WinBUGS}$
[Spiegelhalter et al. (\citeyear{SpiegelhalteretAl2005})]. We used 50,000~iterations with an
initial burn-in of 45,000. Convergence of the generated samples was
assessed using standard tools such as trace plots and ACF plots as well
as Gelman--Rubin diagnostics. The initial values for the fixed
effects parameters were selected arbitrarily. We used two different
MCMC chains (with two different initial values) to be reasonably
confident about convergence. Derivation of the full conditional
posterior distributions of model parameters along with associated
$\tt{WinBUGS}$ code to implement the estimation strategy is provided
in the supplemental article [Bandyopadhyay (\citeyear{Bandyopadhyay2009})].

Our initial model selection was performed using the Deviance
Information Criterion (DIC) of Spiegelhalter et al. (\citeyear{SpiegelhalterEtAl2002}).
DIC reflects the goodness of fit as well as the complexity of the
hierarchical model within the Bayesian paradigm and is considered to
be a Bayesian version of the Akaike Information Criterion (AIC).
It is defined as $\mathit{DIC} = \overline{D} + p_{D}$, where $\overline{D} =
E(D(\bolds{\Theta})|\mathbf{y})$ is the posterior mean of
the deviance and $p_{D}$ is the effective number of parameters in
the model. Spiegelhalter et al. (\citeyear{SpiegelhalterEtAl2002}) showed that $p_{D}$ can be
approximated as $p_{D} = \overline{D} -
D(\hat{\bolds{\Theta}})$, where $\hat{\bolds{\Theta}}$ is
a suitable `plug-in' estimate of $\bolds{\Theta}$, viz. the
posterior mean, or median. The DIC is essentially a single-number
summary (lower is better) of the relative fit between the model and
the `true model' generating the data for the purpose of prediction.
A difference larger than 10 is considered overwhelming evidence in
favor of the better model [Burnham and Anderson (\citeyear{BurnhamAnderson2002})].

After selecting the best model using DIC criterion, we also employed
model validation diagnostics through conditional predictive ordinate
(CPO) statistics [Gelfand, Dey and Chang (\citeyear{GelfandDeyChang1992})] and the associated
`log pseudo-marginal likelihood' (LPML). The CPO is a
cross-validated approach and based on posterior predictive
probability of observed data. For the observed response $y_{ij}$
from youth~$i$ at time $t_{ij}$ with covariate vector
$\mathbf{X}_{ij}$, the CPO statistic for observation $(i,j)$ is
defined as $\mathrm{CPO}_{ij} = f(y_{ij}|{\mathbf{D}}_{(-ij)})
=\int f(y_{ij}|\bolds{\Theta},
\mathbf{X}_{ij})\pi({\bolds{\Theta}}|{\mathbf{D}}_{(-ij)})\,d{\bolds{\Theta}}$, where
$\pi({\bolds{\Theta}}|{\mathbf{D}}_{(-ij)})$ is the
posterior density of parameter vector ${\bolds{\Theta}}$ given
${\mathbf{D}}_{(-ij)}$, the cross-validated data without the
$(i,j)$th observation. Using a harmonic mean approximation result
from\break Gelfand, Dey and Chang (\citeyear{GelfandDeyChang1992}), the $\mathrm{CPO}_{ij}$ can be
easily computed with MCMC samples from the full posterior
$\pi(\bolds{\Theta}|\mathbf{D})$. Typically, the
$\mathrm{CPO}_{ij}$'s behave as Bayesian residuals and are plotted
against any covariate values $x_{ij}$ (or observed $Y_{ij}$'s) to
determine patterns of covariate dependence as well as identify
possible outliers. Larger values of $\mathrm{CPO}_{ij}$ indicate
better support for the model from the observation $y_{ij}$. A
summary measure based on the CPO is the logarithm of the
psuedo-marginal likelihood (LPML) defined as
$\mathrm{LPML}=\sum_{i,j} \log(\mathrm{CPO}_{ij})$, where a
higher value of the LPML means better support of the model from the
observed data.

We observe that Models~1--6 exhibit somewhat similar fit (based on
$\overline{D}$), but the effective parameter count $p_{D}$ greatly
varies. Based on the DIC values, we make a more detailed comparison
between Models~1 and~5 whose DIC values are the lowest, being
821.4 and 832.3 respectively. For these two models, the DIC scores
reflect a considerable trade-off between data fidelity
$(\overline{D})$ and the effective number of parameters $p_{D}$. The
$p_{D}$ value is much lower for Model~1 ($=$63.4) than Model~5
($=$76.6). The DIC value for Model~8 (fixed change-point model) was
somewhat closer ($=$838.8) to Model~5, but for Model~7 (no
change-point model), it was far away ($=$890.6). As described earlier
in Section~\ref{sec2}, a non-Bayesian but convenient procedure such as the
Rao--Scott chi-square test shows a high evidence of association
between the binary indicator of repeat offense and the dichotomized
severity indicator ($p$-value $<$ 0.001). If we omit the binary severe
offense indicator from Model~1, the corresponding interval estimate
of the repeat offense indicator has a slightly tighter 95\% credible
interval ($-$1.041, $-$0.105) than Model~1 at expense of a high DIC
price (859.7).
%
\begin{figure}[b]

\includegraphics{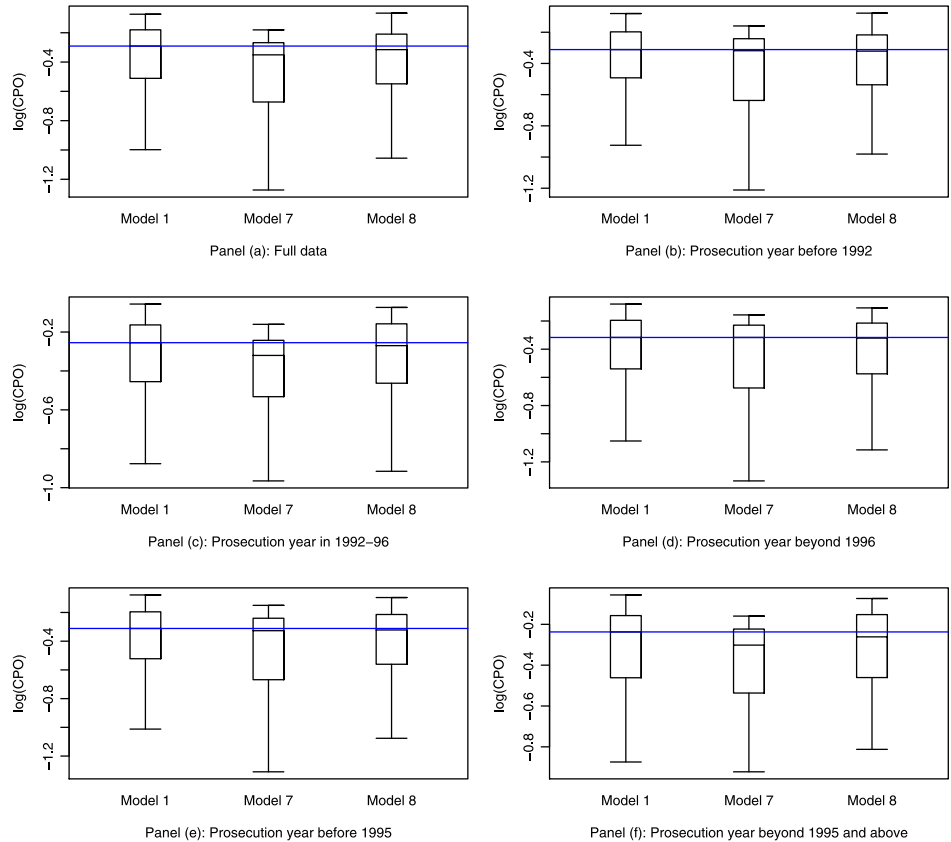}

\caption{Panel box-plots of $\log(\mathrm{CPO})$ for Models 1, 7 and 8
considering different time intervals of prosecution. Larger values
of $\log(\mathrm{CPO})$ indicate more support for the model. The horizontal line
in each panel denotes the median $\log(\mathrm{CPO})$ value for Model~1.}\label{FIG4}
\end{figure}

To assess model validation in terms of predictive performance, we
use the box-plots of $\log(\mathrm{CPO})$ statistics to compare between
Models~1, 7 and 8 in Figure~\ref{FIG4}. The median value of
$\log(\mathrm{CPO})$ for Model~1 is
indicated by the horizontal line in each of the panels. Panel (a)
reveals that on the overall, both Models~1 and~8 (model with fixed
change-point) exhibit significantly better predictive performance
over Model~7 (model with no change-point), though Model~1 performs
marginally better than Model~8. All 3 models have similar fit for
prosecution years before 1992 [panel (b)] as well as beyond 1996
[panel (d)]. For the prosecution years between 1992 and 1996 [panel (c)],
the $\log(\mathrm{CPO})$ values for Model~1 are marginally better than
Model~8 and distinctly better than Model~7. Considering panels (c),
(e) and (f), it is clear that there is strong evidence for a model
with a change-point around 1995. This is clearly demonstrated in
$\log(\mathrm{CPO})$ plots for Models~1 and~8 which includes a change-point
term. The LPML values for Models~1, 7 and 8 are respectively
$-309.78, -350.72$ and $-320.43$, confirming again a
marginally better predictive performance of the unknown change-point
model over the fixed change-point model. While Model~1 includes a
moderate `prior opinion' of 1995 as the change-point year, Model~8
uses the prior view that the change-point is known to be exactly
1995. Model~1 does not prove to be much superior than Model~8
as determined by DIC and predictive performances, yet we choose
Model~1 for further analysis primarily because of the absence of a
common mandate in the literature (before this analysis) of
restricting 1995 as the change-point year for this subset of
juvenile sex offenders with repeated sex offenses.

To determine an overall goodness of fit of Model~1, we also computed
the Bayesian $p$-value [Gelman et al. (\citeyear{GelmanEtAl2004})], which measures the discrepancy
between the data and the model by comparing a summary $\chi^2$
statistic of the posterior predictive distribution to the true
distribution of the data. The summary statistics from the predicted
and observed data are given by $\chi^2(\mathbf{Y},\Theta^g)$ and
$\chi^2(\mathbf{Y}^{\mathit{rep}, g},\Theta^g)$, respectively, where
$\mathbf{Y}^{\mathit{rep},g}$ denote the replicated value of $Y$ from the
posterior predictive distribution of $\Theta$ at the $g$th
iteration of the Gibbs sampler. The Bayesian $p$-value was then
calculated as $P(\chi^2(\mathbf{Y}^{\mathit{rep}, g},\Theta^g)>\chi^2(\mathbf{Y},\Theta^g))$,
that is, the proportion of times $\chi^2(\mathbf{Y}^{\mathit{rep}, g},\Theta^g)$ exceeds
$\chi^2(\mathbf{Y},\Theta^g)$ out of $g=1,\ldots,G$ simulated
draws from the posterior predictive distribution. For Model~1, we
obtain the\break $p$-value to be 0.41 which indicates an overall reasonable
fit, that is, the observed pattern would likely be seen in replications
of the data under the true model.

\section{Results}

Our conclusions of the data analysis were primarily based on Model~
1, which is the best model supported by the data. This model uses
the apriori belief that the most likely year of change in
judicial decision pattern is 1995, however, with a noninformative
prior belief about the effect of the change-point as well as the
attenuation parameter $\phi$. Based on 95\% credible intervals, we
found strong posterior evidence of the effects of several covariates
on the odds of moving forward which includes (a) age at charge, (b)
dichotomized severity of offense, (c)~year of prosecution, (d)
indicator of repeat offense, (e) indicator of whether the time of
offense is after the change-point and (f) interaction between repeat
offense and change-point. In particular, prosecutors were less
likely to move forward on repeated offense than the first-time sex
offenses and less likely to move forward after the change-point. The
strong posterior evidence of the positive interaction effect between
repeat offense and change-point confirms the need to present the
effects of repeat offense charges separately for the time intervals
before and after the change-point.

%
\begin{table}
\caption{Posterior summaries of marginal odds ratios for Model~1.
$T$ denotes unknown time of change-point} \label{TABLE2}
\begin{tabular*}{\textwidth}{@{\extracolsep{\fill}}lccc@{}}
\hline
\textbf{Odds ratio} & \textbf{Mean} & \textbf{Standard deviation} & \textbf{95\% credible intervals} \\
\hline
Age                         & 1.15\phantom{0}  & 0.037 & (1.094, 1.236) \\
Repeat offense before $T$   & 0.604 & 0.163 & (0.308, 0.932) \\
Repeat offense after $T$    & 1.131 & 0.217 & (1.014, 1.639) \\
Severe offense              & 1.506 & 0.280 & (1.049, 2.132) \\
Year of prosecution         & 1.155 & 0.096 & (1.006, 1.373) \\
After change-point effect   & 0.256 & 0.086 & (0.132, 0.483) \\
\hline
\end{tabular*}
\end{table}
%

Table~\ref{TABLE2} presents the posterior estimates together with 95\%
credibility intervals (CI) of `marginal odds ratio' of the
covariates found to be relevant for the prosecutor's decision. The
marginal odds ratio for any particular factor represents the odds
ratio between two randomly selected comparable youths with only a
unit difference in the relevant covariate. We now summarize the
implications of our results. The prosecutors are about 15\% more
likely to move forward on charges for every year increase in age,
about 16\% more likely to move forward for every year increase in
prosecution year. The mean increase in probability of moving forward
is around 51\% for severe offenses as compared to nonsevere
offenses. The odds of prosecution of a charge after the change-point
year will have a 26\% reduction as compared to a charge that
happened before the change-point. Overall, prosecutors were less
likely to move forward on repeat offense cases. However, compared to
a repeat offense charge before the change-point, there is about a
13\% increase in the odds of moving forward for a comparable repeat
offense after the change-point. Interestingly, the magnitude of this
increase in odds may be as low as 1.4\%. The posterior probability
that the change-point occurred in 1995 (maybe due to the
registration policy) for Model~1 is about 61\% (95\% CI between
32\% and 87\%), suggesting strong posterior evidence for 1995 being
the `likely' change-point year compared to any other year between
1992 and 1996. Interestingly, even for the Models~3 and~6 which use the
skeptical prior (c) defined in Section~\ref{sec4}, we find similar strong
evidence that a change-point has occurred in the interval 1992--1996
(results omitted for brevity). In summary, we conclude that there is
strong posterior evidence of the existence of a change-point in
between 1992 and 1996, with the most likely year of change-point being
1995. The posterior estimate of the attenuation parameter $\phi$ is
0.82 (95\% CI between 0.732 and 0.891) and corroborates our apriori belief about existence of moderate degree of heterogeneity
among youths (or heterogeneity due to interactions of different
youths with prosecutors). Interestingly, the posterior intervals of
$\phi$ are very close for all six models, indicating that the data
supports strongly our skeptical prior belief about $\phi$. For Model~
5, the 95\% posterior CI of severity indicator effect barely
covers zero, indicating lack of overwhelming evidence of the
existence of severity effect when we are skeptical about 1995 being
the change-point year.
%
\begin{figure}

\includegraphics{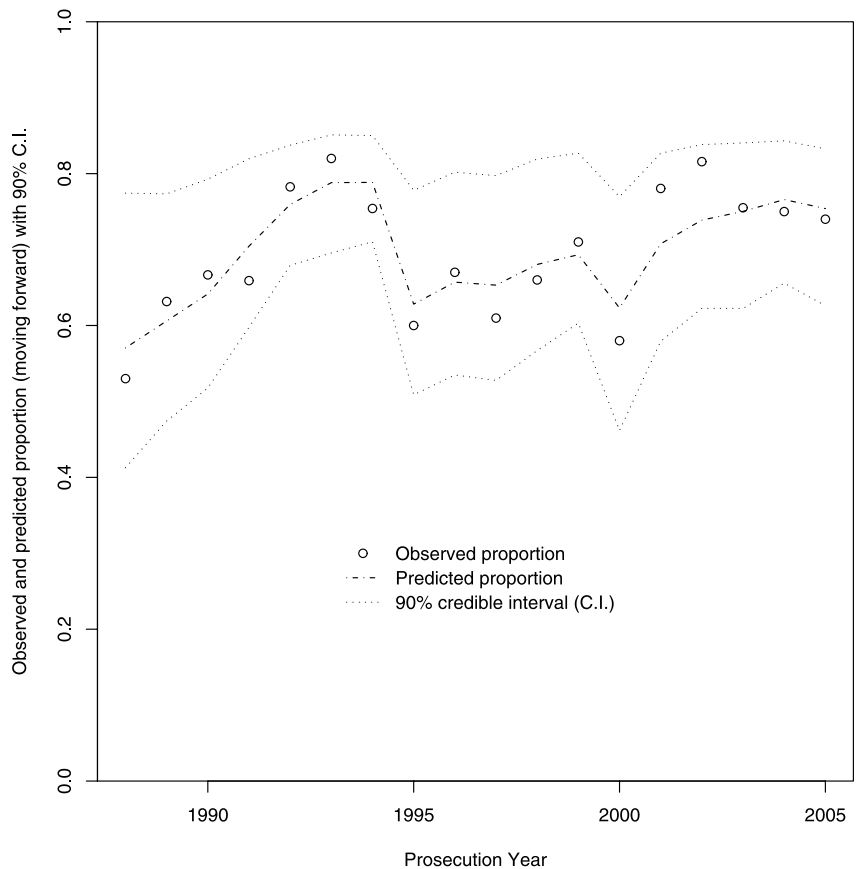}

\caption{Plots of observed and fitted posterior proportion $p$ of
prosecutor's moving forward versus prosecution years along with 90\%
credible intervals for Model~1.} \label{FIG5}
\end{figure}
Figure~\ref{FIG5} plots the observed and predicted proportions of
prosecutor's moving forward along with the 90\% CI for the
prosecution years between 1988 and 2005. Our semi-continuous
(change-point) model clearly captures the observed trend including
the substantial reductions in 1994--1995 along with the effect around
1999--2000. All the observed proportions are found to lie within the
90\% CI of the predicted ones. Our data analysis results confirm
that even using the most skeptical prior belief, there is enough
posterior evidence to support the effect of most of the variables on
prosecutors' decision making pattern over time. The posterior
evidence is, however, inconclusive for the effect of the 1999
internet-based notification policy as well as for the linear and
quadratic terms confirming that the change-point is abrupt rather
than gradual. There was also no evidence of any gradual and
prevailing effects of this change-point from the predicted
proportions in Figure~\ref{FIG5}.
%
\begin{figure}

\includegraphics{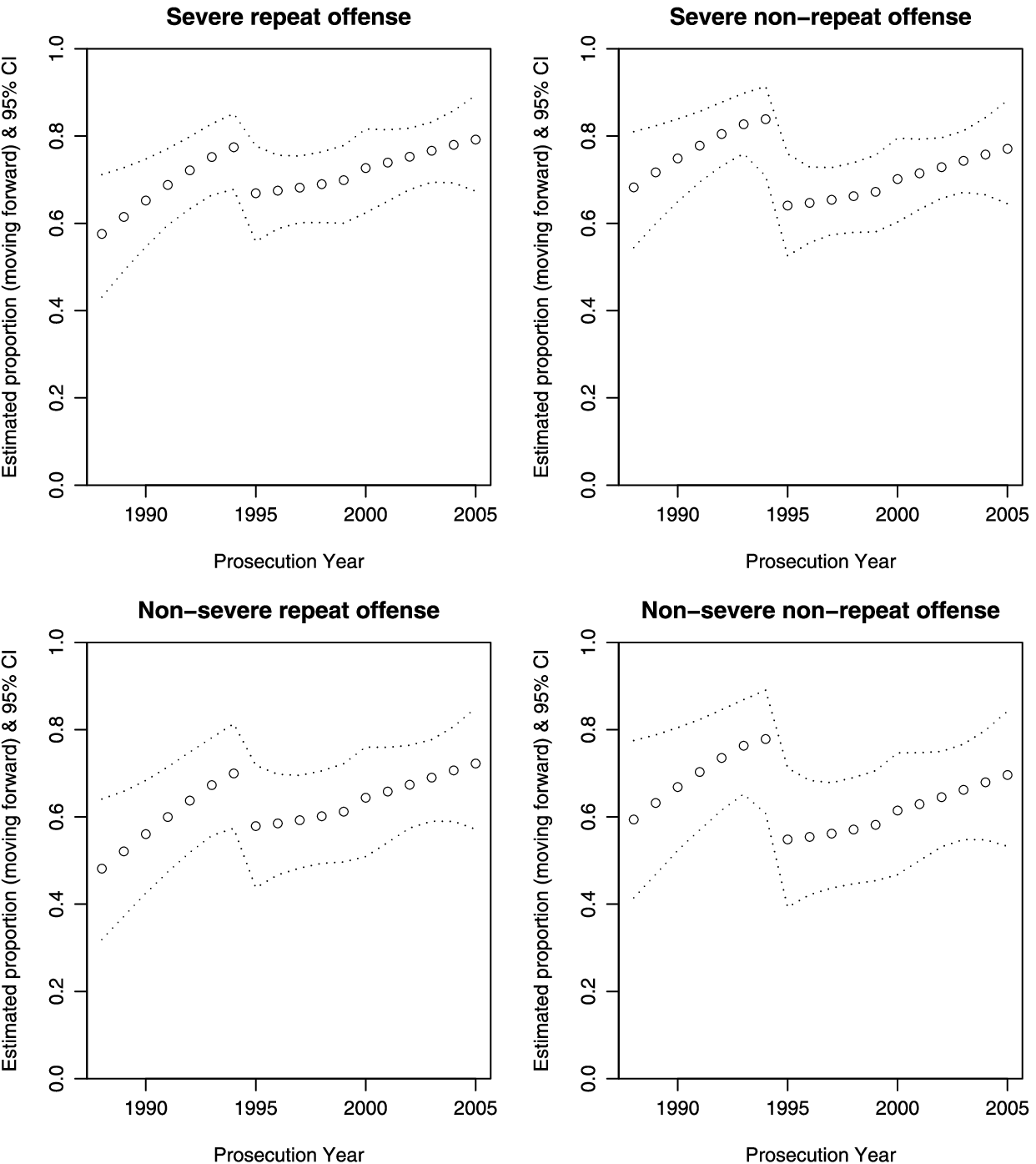}

\caption{Plots of estimated posterior proportion $p$ of prosecutor's
moving forward on juvenile sex-offenders with median ($=$14.6 years)
age along with 95\% credible intervals for Model~1.} \label{FIG6}
\end{figure}

Figure~\ref{FIG6} depicts plots of posterior predictive probabilities along
with 95\% CI estimates of prosecutor's moving forward for
prosecution years using various combinations of median (14.6 years)
age, severe/nonsevere and first/repeat offenses. All of these plots
(Figures~\ref{FIG5} and~\ref{FIG6}) reflect the apparent posterior evidence of a
change-point around 1995. In 1995, a nonrepeat offense had a larger
magnitude of the decrease in the posterior probability from the
previous year compared to the corresponding decrease for a repeated
offense. The decrease in magnitude of the probability in 1995 is
largest for first-time nonsevere offenses. For the plots of severe
and nonsevere repeat offenses in Figure~\ref{FIG6}, there does appear to be
a reduction in the probability of prosecutors moving forward on
repeat offenses starting in~1995 (the year registration was
implemented) and this effect decreases over time. The 95\% CI seem
to be wider for the nonsevere offenses than the severe offenses,
indicating more posterior uncertainty about these probabilities over
time for nonsevere offenses compared to that of severe offenses.

\section{Concluding remarks and policy implications}
Using data from South Carolina juvenile male repeat sex offenders,
this study examined how during 1992--1996, the change in pattern of
prosecutor's decision substantively altered the consequences faced
by these youths. Specifically, from 1995 through the present,
juveniles adjudicated as minors for certain sexual offenses have
faced lifetime registration and many of these youths have been
subjected to broad community notification via inclusion in South
Carolina's Internet-based sex offender registry site. Results from
an earlier GEE-based analysis [Letourneau et al. (\citeyear{LetourneauEtAl2009})] using a much
bigger sample (cohort) also suggested that prosecutors altered their
behavior specifically in response to the 1995 legislation, such that
they became less likely to move forward on serious sexual offense
cases. However, this previous study included an overwhelming
proportion of first-time (or single) sexual offenses and was based
on the strong apriori assumption that the change-point year
was known to be 1995. The present study sought to expand on these
earlier findings by focusing on the prosecution of youths charged
with repeated sexual offenses. The novelty of this extensive
analysis lies in utilizing the Bayesian paradigm for making useful
and interpretable conclusions from a complex project involving
multiple research questions and different prior opinions. The
analytic strategy (using Bridge random effects in a longitudinal
model with unknown change-points) permitted addressing and
evaluating the heterogeneity of the youths and determining
attenuation effects after adjusting for this heterogeneity
simultaneously.

In many respects, results from our analysis further the findings
from our previous research. First, there was strong support for a
significant change-point occurring within the 1992--1996 time frame,
and particularly for a 1995 change-point. Second, as we have
previously found, prosecutors were more likely to move forward on
older defendants, defendants with more (vs. fewer) prior
adjudications and defendants charged with more severe sexual
offenses. Two compelling results suggest that applying lifetime sex
offender registration requirements to juvenile offenders altered
prosecutor behavior. First, prosecutors were less likely to move
forward on sex offense cases after than before the change-point,
with particularly strong evidence of a 1995 change-point, the year
registration was implemented. Second, there also was evidence that
prosecutors were generally less likely to move forward on repeat sex
offense cases than on initial sex offense cases. As depicted in
Figure~\ref{FIG6}, the reduction of probability of moving forward appears to
have been strongest around 1995 for both repeat and nonrepeat
offenses. After a significant drop in the odds of moving forward on
repeat cases before the change-point, prosecutors became somewhat
more likely to do so over time. Thus, the chilling effect of
lifetime registration on the prosecution of serious repeat sexual
offenses might be declining.

Policy implications of these findings are necessarily limited by the
need to replicate results with data from other states/population. At
minimum, however, it appears safe to state that SC's experiment with
the lifetime registration of juvenile sexual offenders is having
unintended effects of reducing the probability of prosecution of
these youths, which in turn may adversely affect community safety
via reduced supervision and treatment of juvenile sex offenders. In
light of concerns about latent consequences of public registration
to juvenile offenders [Chaffin (\citeyear{Chaffin2008});
Trivits and Reppucci (\citeyear{TrivitsReppucci2002})]
and the typically low sexual recidivism risk posed by juvenile
sexual offenders [Fortune and Lambie (\citeyear{FortuneLambie2006})], results from our
studies suggest that state and federal registration policies could
be revised without increasing the risk of harm to community members.
In particular, policies in which long term public registration
requirements are trigged solely on juvenile adjudication offense
(and not other indicators of recidivism risk) should be targeted for
modification. When prosecutors believe that only the most severe and
highest risk offenders will face long term and/or public
registration, they may be less likely to alter their judicial
behavior. Three specific modifications may achieve this aim. First,
to reduce the threat of harmful latent consequences to youth,
offenders adjudicated as minors should not be subjected to broad
community notification requirements (e.g., should not be included on
Internet-based registry websites). Second, to ensure that
registration targets high risk offenders, registration requirements
should be based on comprehensive risk assessments, as is currently
the case in several states. Third, the duration of registration
requirements should reflect developmental differences between
juvenile and adult offenders (and between younger and older
juveniles). For example, as is the case with duration of probation,
registration requirements could end with the offender reaching the
age of majority in his or her state in the absence of subsequent
sexual or violent offenses. These changes might permit judicial
decision makers to have greater confidence that youth targeted by
registration policies are, indeed, deserving of the consequences
that attend these policies and such confidence should reduce the
unintended effects of registration policies on judicial decision
making.

\section*{Acknowledgments}
We are thankful to
the Editor and the Associate Editor whose constructive comments led
to a significant improvement in the presentation.

\begin{supplement}[id=suppA]
\sname{Supplement}
\stitle{Posterior computations and
code for \textit{Changing approaches of prosecutors toward juvenile
repeated sex-offenders: A Bayesian evaluation}}
\slink[doi]{10.1214/09-AOAS295SUPP}
\slink[url]{http://lib.stat.cmu.edu/aoas/295/supplement.pdf}
\sdatatype{.pdf}
\sdescription{The web supplement provides derivation of the
conditional posterior distributions as well as the associated
$\tt{WinBUGS}$ code for the analysis.}
\end{supplement}

\printaddresses


\begin{thebibliography}{99}

\bibitem[\protect\citeauthoryear{}{2009}]{Bandyopadhyay2009}
\textsc{Bandyopadhyay, D.} (2009).
Supplement to ``Changing approaches of prosecutors towards juvenile
repeated sex-offenders: A~Bayesian evaluation''
DOI: \href{http://dx.doi.org/10.1214/09-AOAS295SUPP}{10.1214/09-AOAS295SUPP}.

\bibitem[\protect\citeauthoryear{}{2006}]{BarrettKatsiyanisZhang2006}
\textsc{Barrett, D.~E., Katsiyanis, A.} and \textsc{Zhang, D.}
(2006). Predictors
of offense severity, prosecution, incarceration and repeat
violations for adoloscent male and female offenders.
\textit{Journal of Child and Family Studies} \textbf{15} 709--719.

%

\bibitem[\protect\citeauthoryear{Bumby, Talbot and Carter}{2009}]{BumbyTalbotCarter2008}
\textsc{Bumby, K. M., Talbot, T. B.} and \textsc{Carter, M. M.}
(2009). Sex
offender reentry: Facilitating public safety through successful
transition and community reintegration.
\textit{Criminal Justice and Behavior}. To appear.

\bibitem[\protect\citeauthoryear{}{2002}]{BurnhamAnderson2002}
\textsc{Burnham, K.~P.} and \textsc{Anderson, D.~R.} (2002).
\textit{Model~Selection and Multivariate Inference: A~Practical
Information-Theoretic Approach}, 2nd ed. Springer, New York.
\MR{1919620}


\bibitem[\protect\citeauthoryear{}{2002}]{Caldwell2002}
\textsc{Caldwell}, M. F. (2002).
What we do not know about juvenile sexual reoffense risk.
\textit{Child Maltreatment} \textbf{7} 291--302.

\bibitem[\protect\citeauthoryear{}{2000}]{CarlinLouis2000}
\textsc{Carlin, B.~P.} and \textsc{Louis, T.~A.} (2000).
\textit{Bayes and Empirical Bayes Methods for Data Analysis}, 2nd ed.
Chapman and
Hall/CRC Press, Boca Raton, FL.
\MR{1427749}

\bibitem[\protect\citeauthoryear{}{2008}]{Chaffin2008}
\textsc{Chaffin, M.} (2008). Our minds are made up: Don't confuse
us with the facts.
\textit{Child Maltreatment [Special Issue: Children with Sexual
Behavior Problems]} \textbf{13} 110--121.

\bibitem[\protect\citeauthoryear{}{1998}]{Chaiken1998}
\textsc{Chaiken, J.~M.} (1998). Introduction paper presented at the
Bureau of Justics Statistics National Conference on Sex Offender
registries. Available at \url{http://www.ojp.usdoj.gov/bjs/pub/ascii/ncsor.txt}
(retrieved February~21, 2003).


\bibitem[\protect\citeauthoryear{}{2001}]{EdwardsHensley2001}
\textsc{Edwards, W.} and \textsc{Hensley, C.} (2001). Contextualizing sex
offender management legislation and policy: Evaluating the problem
of latent consequences in community notification laws.
\textit{International Journal of Offender Theraphy and Comparative
Criminology} \textbf{45} 83--101.

\bibitem[\protect\citeauthoryear{}{2006}]{FortuneLambie2006}
\textsc{Fortune, C.-A.} and \textsc{Lambie, I.} (2006). Sexually abusive
youth: A~review of recidivism studies and methodological issues for
future research.
\textit{Clinical Psychology Review} \textbf{26} 1078--1095.

\bibitem[\protect\citeauthoryear{}{2003}]{Garfinkle2003}
\textsc{Garfinkle, E.} (2003). Coming of age in America: The
misapplication of sex-offender registration and
community-notification laws to juveniles.
\textit{California Law Review} \textbf{91} 163--208.

\bibitem[\protect\citeauthoryear{}{1992}]{GelfandDeyChang1992}
\textsc{Gelfand, A.~E., Dey, D.~K.} and \textsc{Chang, H.} (1992). Model
determination using predictive distributions with implementation via
sampling-based methods (with discussion).
In \textit{Bayesian Statistics~4}
(J.~M. Bernardo, J.~O. Berger, A.~P. David and A.~F.~M.~Smith, eds.) 147--167
Oxford Univ. Press, Oxford.
\MR{1380275}

\bibitem[\protect\citeauthoryear{}{1990}]{GelfandSmith1990}
\textsc{Gelfand, A.} and \textsc{Smith, A. F. M.} (1990).
Sampling based approaches to calculating marginal densities.
\textit{J. Amer. Statist. Assoc.} \textbf{85} 398--409.
\MR{1141740}

\bibitem[\protect\citeauthoryear{}{2004}]{GelmanEtAl2004}
\textsc{Gelman, A., Carlin, J.~B., Stern, H.} and \textsc{Rubin, D.} (2004).
\textit{Bayesian Data Analysis}, 2nd ed.
Chapman and Hall/CRC, New York.
\MR{2027492}

\bibitem[\protect\citeauthoryear{}{1992}]{GilksWild1992}
\textsc{Gilks, W.~R.} and \textsc{Wild, P.} (1992). Adaptive rejection
sampling for Gibbs sampling.
\textit{Appl. Statist.} \textbf{41} 337--348.

%

\bibitem[\protect\citeauthoryear{}{2003}]{Howell2003}
\textsc{Howell, J. C.} (2003).
\textit{Preventing and Reducing Juvenile Delinquency: A~Comprehensive
Framework}.
Sage Publications, Thousand Oaks, CA.

\bibitem[\protect\citeauthoryear{}{2007}]{Kaban2007}
\textsc{Kaban, A.} (2007). On Bayesian classification with Laplace priors.
\textit{Pattern Recognition Letters} \textbf{28} 1271--1282.

%

\bibitem[\protect\citeauthoryear{}{2005}]{LaFond2005}
\textsc{LaFond, J. Q.} (2005).
\textit{Preventing Sexual Violence: How Society Should Cope With Sex Offenders}.
Am. Psychol. Assoc., Washington, DC.

%

\bibitem[\protect\citeauthoryear{}{2005}]{LetourneauMiner2005}
\textsc{Letourneau, E.} and \textsc{Miner, M.~H.} (2005). Juvenile sex
offenders: A~case against the legal and clinical status quo.
\textit{Sexual Abuse: Journal of Research and Treatment} \textbf{17} 293--312.

\bibitem[\protect\citeauthoryear{}{2009}]{LetourneauEtAl2009}
\textsc{Letourneau, E., Bandyopadhyay, D., Sinha, D.} and \textsc{Armstrong,
K.} (2009). Effects of sex offender registration policies on
juvenile justice decision making.
\textit{Sexual Abuse: A~Journal of Research and Treatment} \textbf
{21} 149--165.

\bibitem[\protect\citeauthoryear{}{2005}]{LevensonCotter2005}
\textsc{Levenson, J.~S.} and \textsc{Cotter, L.~P.} (2005).
The effect of Megans Law on sex offender reintegration.
\textit{Journal of Contemporary Criminal Justice} \textbf{21} 49--66.

\bibitem[\protect\citeauthoryear{}{2005}]{McManus2005}
\textsc{McManus, R.} (2005).
\textit{South Carolina Criminal and Juvenile Justice Trends: 2004}.
South Carolina Dept. Public Safety, Columbia, SC.

%
%
%
%
\bibitem[\protect\citeauthoryear{}{2003}]{WangLouis2003}
\textsc{Wang, Z.} and \textsc{Louis, T. A.} (2003). Matching
conditional and
marginal shapes in binary mixed-effects models using a bridge
distribution function.
\textit{Biometrika} \textbf{90} 765--775.
\MR{2024756}

\bibitem[\protect\citeauthoryear{}{2005}]{SpiegelhalteretAl2005}
\textsc{Spiegelhalter, D., Thomas, A., Best, N.} and \textsc{Lunn, D.} (2005).
WinBUGS user manual, version~1.4.2, MRC Biostatistics Unit,
Institute of Public Health and Dept. Epidemiology and Public Health,
Imperial College School of Medicine.
Available at \url{http://www.mrc-bsu.cam.ac.uk/bugs}.

\bibitem[\protect\citeauthoryear{}{2002}]{SpiegelhalterEtAl2002}
\textsc{Spiegelhalter, D.~J., Best, N.~G., Carlin, B.~P.} and \textsc{van der
Linde, A.} (2002). Bayesian measures of model complexity and fit
(with discussion).
\textit{J.~Roy. Statist. Soc. Ser.~B} \textbf{64}
583--639.
\MR{1979380}

\bibitem[\protect\citeauthoryear{}{2004}]{TerryFurlong2004}
\textsc{Terry, K. J.} and \textsc{Furlong, J. S.} (2004).
\textit{Sex Offender Registration and Community Notification:
A~`Megan's Law' Sourcebook}.
Civic Research Institute, Kingston, NJ.

\bibitem[\protect\citeauthoryear{}{2005}]{Tewksbury2005}
\textsc{Tewksbury, R.} (2005). Collateral consequences of sex
offender registration.
\textit{Journal of Contemporary Criminal Justice} \textbf{21} 67--82.

\bibitem[\protect\citeauthoryear{}{2002}]{TrivitsReppucci2002}
\textsc{Trivits, L. C.} and \textsc{Reppucci, N. D.} (2002).
Application of
Megan's Law to juveniles.
\textit{American Psychologist} \textbf{57} 690--704.

\bibitem[\protect\citeauthoryear{}{1986}]{ZegerLiang1986}
\textsc{Zeger, S. L.} and \textsc{Liang, K.-Y.} (1986).
Longitudinal data analysis for discrete and continuous outcomes.
\textit{Biometrics} \textbf{42} 121--130.

\bibitem[\protect\citeauthoryear{}{2006}]{Zevitz2006}
\textsc{Zevitz, R. G.} (2006). Sex offender community notification:
Its role in recidivism and offender reintegration.
\textit{Criminal Justice Studies} \textbf{19} 193--208.

\bibitem[\protect\citeauthoryear{}{2004}]{Zimring2004}
\textsc{Zimring, F. E.} (2004). \textit{An American Travesty: Legal
Responses to Adolescent Sexual Offending}.
Chicago Univ. Press.

\bibitem[\protect\citeauthoryear{}{2007}]{ZimringPiqueroJennings2007}
\textsc{Zimring, F.~E., Piquero, A.~R.} and \textsc{Jennings, W.~G.} (2007).
Sexual delinquency in racine: Does early sex offending predict later
sex offending in youth and young adulthood?
\textit{Criminology and Public Policy} \textbf{6} 507--534.
\end{thebibliography}
\end{document}